\documentclass[%
 reprint, amsmath,amssymb,
 aps,
pra,groupedaddress,twocolumn,floatfix
]{revtex4-2}

\usepackage{graphicx}
\usepackage{dcolumn}
\usepackage{bm}
\usepackage{mathtools}
\usepackage{upgreek}

\DeclareMathOperator{\Tr}{Tr}

\DeclarePairedDelimiterX{\abs}[1]{\vert}{\vert}{#1}
\DeclarePairedDelimiterX{\norm}[1]{\lVert}{\rVert}{#1}
\DeclarePairedDelimiterX{\expval}[1]{\langle}{\rangle}{#1}
\DeclarePairedDelimiterX{\ket}[1]{\vert}{\rangle}{#1}
\DeclarePairedDelimiterX{\bra}[1]{\langle}{\vert}{#1}
\DeclarePairedDelimiterX{\innerproduct}[2]{\langle}{\rangle}{#1\delimsize\vert\mathopen{}#2}
\DeclarePairedDelimiterX{\outerproduct}[2]{\vert}{\vert}{#1\delimsize\rangle\!\delimsize\langle\mathopen{}#2}
\DeclarePairedDelimiterX{\mel}[3]{\langle}{\rangle}%
{#1\delimsize\vert\mathopen{}#2\delimsize\vert\mathopen{}#3}

\begin{document}

\preprint{APS/123-QED}

\title{Demonstration of Quantum-Limited Discrimination of Multi-Copy Pure versus Mixed States}

\author{Arunkumar Jagannathan$^{1}$, Michael Grace$^{2}$, Olivia Brasher$^{3}$, Jeffrey H. Shapiro$^{4}$, Saikat Guha$^{2}$, and Jonathan L. Habif$^{1,3}$}
\affiliation{$^{1}$ Information Sciences Institute, University of Southern California,Waltham, MA 02451}
\affiliation{$^{2}$ James C. Wyant College of Optical Sciences, University of Arizona, Tucson, AZ 85721}
\affiliation{$^{3}$ Dept. of Elec. \& Comp. Engineering, University of Southern California, Los Angeles, CA 90089}
\affiliation{$^{4}$ Research Laboratory of Electronics, Massachusetts Institute of Technology, Cambridge, MA 02139}

\date{\today}

\begin{abstract}
We demonstrate an optical receiver that achieves the quantum Chernoff bound for discriminating coherent states from thermal states in the multi-copy scenario. In contrast, we find that repeated use of the receiver approaching the Helstrom bound for single-copy measurement is sub-optimal in this multi-copy case. Furthermore, for a large class of multi-copy discrimination tasks between a pure and a mixed state, we prove that any Helstrom-bound achieving single-copy receiver is sub-optimal by a factor of at least two in error-probability exponent compared to the multi-copy quantum Chernoff bound.  This behavior has a classical analog in the performance gap between soft-decision and hard-decision receivers for detecting a multi-copy signal embedded in white Gaussian noise.
\end{abstract}

\maketitle

\section{Introduction}
Helstrom~\cite{Helstrom1964} launched the field of quantum hypothesis testing by deriving the optimum measurement operator for minimizing the error probability of single-copy discrimination between optical states.  His measurement's error probability---known as the Helstrom bound---is the gold standard to which all other receivers aspire.  His work has led to a wealth of research devoted to quantum-limited single-copy discrimination of the symbols received in photon-starved laser communications.  Nearly the entirety of these investigations have focused on discriminating between quantum pure states---typically the coherent states produced by ideal lasers---comprising a symbol constellation.  Receivers that approach or achieve the Helstrom bound have been proposed and demonstrated for coherent-state constellations of size $N = 2$ \cite{wittmann2008demonstration,dolinar1973,Geremia,Kennedy,PhysRevLett.121.023603} and $N > 2$ \cite{guha2011approaching,chen2012optical,becerra2013experimental,izumi2013quantum,ferdinand2017multi}, including constellations suffering from phase noise~\cite{PhysRevResearch.2.023384}.  These quantum measurements all follow a general architecture of coherent-state displacement followed by photon counting.  Interestingly, this same quantum measurement prescription was proposed in \cite{PhysRevA.89.032318} for achieving optimal discrimination for \emph{multi-copy} quantum states, proving that such a receiver can achieve the quantum Chernoff bound's (QCB's) error-probability exponent \cite{Audenaert}.  

Recent investigations have addressed discrimination tasks involving mixed states \cite{habif2018quantum,lu2018quantum,zhuang2017optimum,cohen2019thresholded,cohen2021towards}. Experimental work has demonstrated techniques for improved discrimination between single-copy pure and mixed quantum states \cite{you2020identification}, including one that approached the Helstrom bound \cite{habif2021quantum}.  Little work, experimental or theoretical, has been devoted to the quantum limits on multi-copy pure versus mixed-state discrimination.

In this paper we present the first experimental demonstration of a multi-copy receiver for discriminating quantum optical states at the QCB. Specifically, over a broad range of average photon numbers we show that the Kennedy receiver \cite{Kennedy}---originally proposed for binary phase-shift keyed laser communication---can do multi-copy discrimination between coherent states and thermal states at the QCB.  Moreover, we show experimentally that repeated use of the single-copy receiver that approaches the Helstrom bound for coherent-state versus thermal-state discrimination is strictly sub-optimal for the multi-copy case.   We also prove, theoretically, that this multi-copy sub-optimality of the single-copy Helstrom-bound achieving measurement holds for all discrimination tasks between pure and mixed states obeying a broadly applicable symmetry condition.  Furthermore, we exhibit an architecture that realizes QCB-achieving multi-copy reception for discriminating between a squeezed state and a mixed state, and we show that development of a vacuum-or-not implementation for arbitrary pure states would enable the same to be done for the general case of multi-copy pure-state versus mixed-state reception.  
 
\section{Discrimination task}
The multi-copy ($M$-ary) discrimination task we will address is as follows.  Under the coherent-state hypothesis, the density operator for the received light is $\rho_R^{\otimes M} = \rho_{\rm coh}^{\otimes M}$, where $\rho_{\rm coh} = |\alpha\rangle\langle \alpha|$ is the single-copy coherent-state density operator with average photon number $\bar{n} = |\alpha|^2$.  Under the thermal-state hypothesis, the density operator for the received light is $\rho_R^{\otimes M} = \rho_{\rm th}^{\otimes M}$, where $\rho_{\rm th} = \sum_{n=0}^\infty\frac{\bar{n}^{n}}{(\bar{n}+1)^{n+1}} |n\rangle \langle n|$ is the single-copy thermal-state density operator with average photon number $\bar{n}$ and $\{|n\rangle\}$ is the photon-number basis.  We take a Bayesian approach in which the two hypotheses are equally likely and we quantify receiver performance using the error probability $P^{(M)}_{\epsilon}=[P(\mbox{decide }\rho_{\rm coh}^{\otimes M}\vert \rho_{\rm th}^{\otimes M}\mbox{ true})+P(\mbox{decide }\rho_{\rm th}^{\otimes M}\vert\rho_{\rm coh}^{\otimes M}\mbox{ true})]/2$.  

Our interest in the multi-copy case stems from photon-starved operation, i.e., $\bar{n}$ is sufficiently low that the single-copy error probability, $P^{(1)}_{\epsilon}$ is close to 1/2.  
The multi-copy error probability for any particular measurement satisfies the Chernoff bound, 
$P^{(M)}_{\epsilon} \leq \exp\left(-M\xi^{\rm Meas}\right)/2$, 
where $\xi^{\rm Meas}$ is that measurement's Chernoff exponent \cite{VanTrees2013}.  The Chernoff bound is known to be exponentially tight, i.e., $-\lim_{M\rightarrow \infty}[\ln(2P^{(M)}_\epsilon)/M] = \xi^{\rm Meas}$.  It follows that sufficiently large $M$ provides acceptable performance even when $P^{(1)}_{\epsilon}$ is close to 1/2.  More importantly, the quantum Chernoff bound $\xi^{\rm QCB}\geq\xi^{\rm Meas}$ upper bounds the Chernoff exponent for \emph{all} physically possible measurements by the quantum Chernoff exponent $\xi^{\rm QCB}\equiv -\ln\!\left[\min_{0\leq s \leq 1} \Tr(\rho_{\rm coh}^{s}\rho_{\rm th}^{1-s})\right]$, which is the ultimate benchmark for this problem \cite{Audenaert,Nussbaum2009}.

\begin{figure}
    \centering
    \includegraphics[width = \columnwidth]{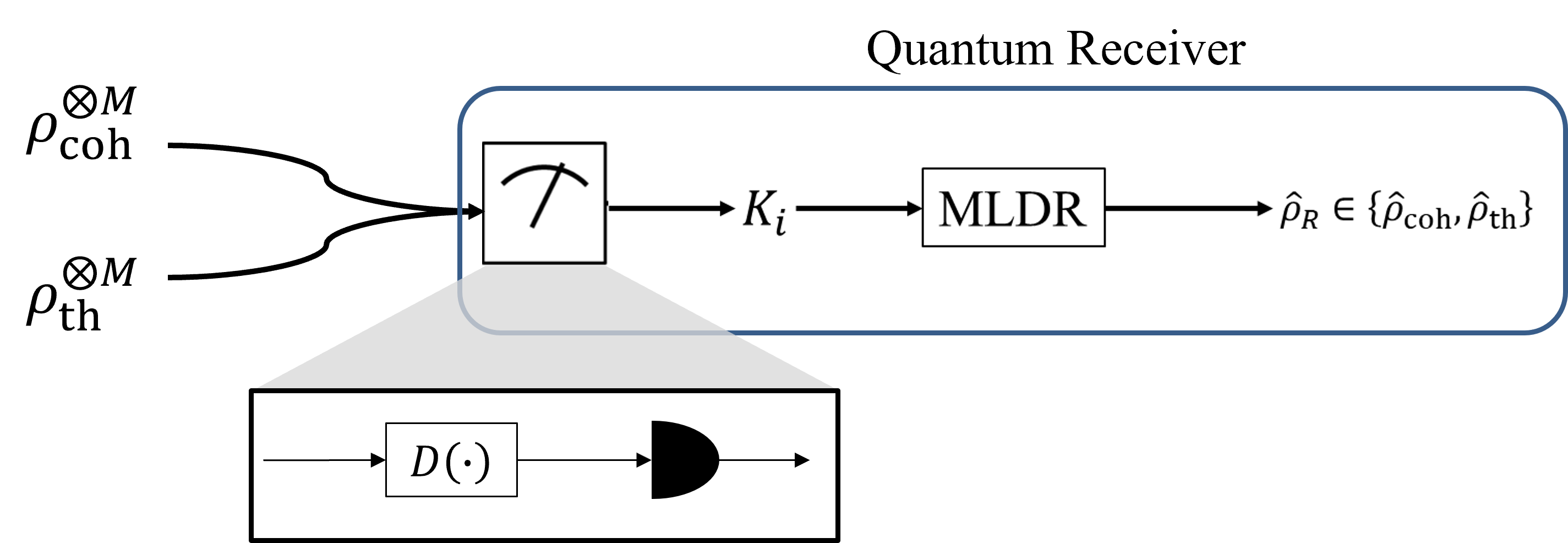}
    \caption{Schematic of multi-copy quantum discrimination.  Two multi-copy sources are equally likely to illuminate the receiver.  Copy-by-copy quantum measurements are made on the $M$ arriving states and the measurement results ($K_{i}$) are employed in a maximum likelihood decision rule (MLDR) to decide which source was present.}
    \label{fig:ProblemStatement}
\end{figure}

\section{Receivers}
Fig.~\ref{fig:ProblemStatement} shows the multi-copy receiver's architecture, viz.,  a quantum measurement followed by the maximum likelihood decision rule (MLDR) to decide which state was received.  In general, achieving QCB performance from this architecture requires optimal joint measurement over all $M$ copies. In practice, finding and implementing that optimal joint measurement can be very difficult, cf.~\cite{zhuang2017optimum}.  However, joint measurement is not required to achieve the QCB in our coherent-state versus thermal-state scenario, because one of those states is pure. Indeed, in pure-state versus mixed-state multi-copy discrimination projecting all $M$ copies of the received state onto the pure state, followed by post-processing of each copy's measurement outcome achieves the QCB~\cite{kargin2005chernoff}. This procedure identifies the pure state with certainty, providing a sufficient statistic for the hypothesis test that maximizes $\xi^{\rm Meas}$ despite its \emph{not} minimizing the single-copy error probability.  The resulting quantum Chernoff exponent for pure-state versus mixed-state discrimination is $\xi^{\rm QCB}=-\ln\!\big[F(\rho_1,\rho_2)\big]$, where $F(\rho_1,\rho_2)\equiv\Tr\!\big[\sqrt{\sqrt{\rho_1}\rho_2\sqrt{\rho_1}}\big]^2$ is the quantum fidelity between the two states' density operators~\cite{kargin2005chernoff}.

Our experiments used the Kennedy receiver~\cite{Kennedy} and the generalized Kennedy receiver~\cite{wittmann2008demonstration}.   Both perform copy-by-copy coherent-state displacement on their incoming light followed by copy-by-copy photon-number resolving detection.  
The $M$ outputs from the photon counter then undergo multi-copy MLDR processing to decide between the two hypotheses.  The difference between the Kennedy and generalized Kennedy receivers lies in their displacements:  the Kennedy receiver displaces $|\alpha\rangle$ to the vacuum $|0\rangle$, whereas the generalized Kennedy (GK) receiver displaces $|\alpha\rangle$ 
by a pre-computed amount that minimizes $P_{\epsilon}^{(1)}$. We compare these against a direct-detection (DD) receiver, in which copy-by-copy photon-number resolving detection and an MLDR are performed \emph{without} coherent-state displacement.

We previously presented~\cite{habif2021quantum} an experimental demonstration of the Kennedy, GK, and DD receivers for single-copy discrimination between coherent-state and thermal-state light of the same average photon number. The experiments showed that the GK receiver closely approaches the Helstrom bound, the Kennedy receiver is sub-optimal, and both greatly outperform the DD receiver. For multi-copy discrimination, our work in this paper shows, both experimentally and analytically, that:  (1) the Kennedy receiver exactly attains the QCB, as expected~\cite{habif2018quantum,kargin2005chernoff}: (2) the GK receiver is decidedly not optimal by a provable factor; and (3) both greatly outperform the DD receiver.  
\begin{figure*}[tb]
	\includegraphics[width = \textwidth]{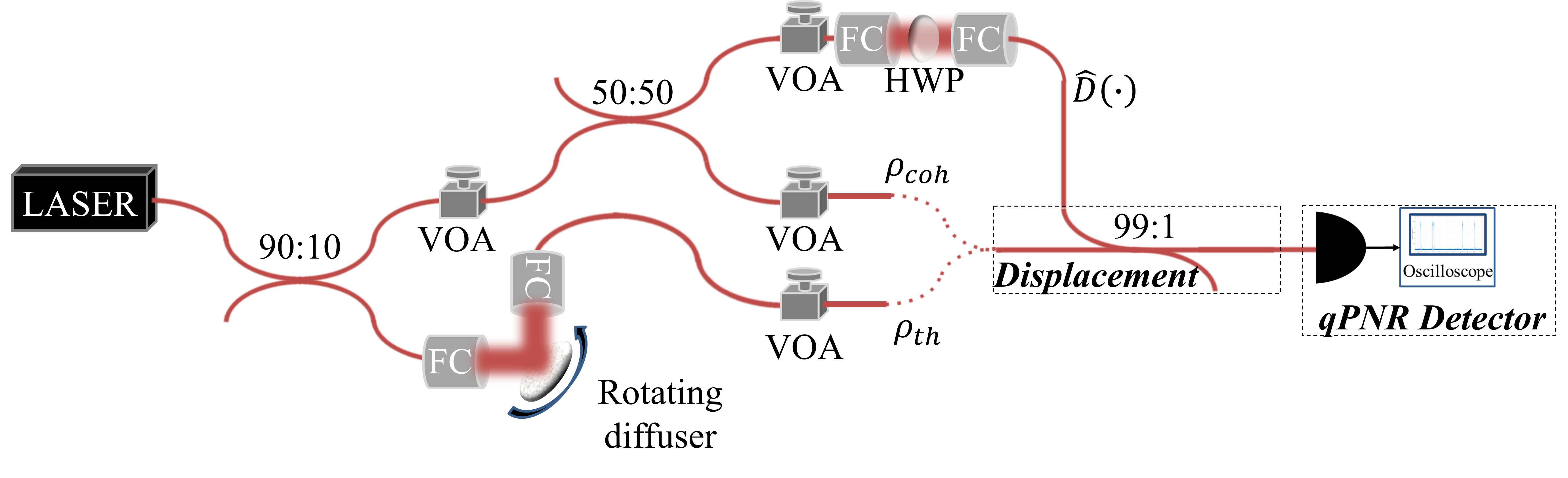}
	\caption{Experimental setup for multi-copy discrimination between coherent and thermal states. VOA: variable optical attenuator. HWP: Half-wave plate. $\hat{D}(\cdot)$: displacement operator. qPNR: quasi photon-number resolving detector.    	\label{fig:ExpSetup}
	}
\end{figure*}

\section{Experiments}
\subsection{Setup}
Figure~\ref{fig:ExpSetup} shows our experimental setup. A shot-noise limited 780\,nm laser (Toptica DL100 Pro) generates continuous-wave light approximating a coherent-state source for our experiments.  Reflecting laser light from a rotating diffuser and collecting a portion of the resulting speckle pattern with a single-mode optical fiber generates a time-varying single spatiotemporal-polarization mode with $\sim$1\,$\upmu$s coherence time.  We verified that light from the fiber had the single-mode thermal state's Bose-Einstein photon-counting statistics over $\upmu$\,s-duration intervals~\cite{habif2021quantum}, as well as its $g^{(2)}(0) \sim 2$ second-order correlation function at zero lag. 

The coherent state and thermal state photon fluxes were independently adjustable, using variable optical attenuators, to keep their average photon numbers equal to $\bar{n}_{\rm S}$ as that value was varied.  The Kennedy and GK receivers' displacement operations were performed by harvesting a portion of the initial laser beam and mixing it with the signal state passing through a highly transmissive (99:1) beam splitter.  Following the displacement, direct detection was performed using the quasi photon-number resolving (qPNR) detector (described in Appendix~\ref{AppA} and Ref.~\cite{habif2021quantum}) consisting of a photon-counting Si avalanche photodiode (Perkin Elmer SPCM-AQR-14) and an oscilloscope for counting individual photon detections.  

The Kennedy, GK, and DD receivers were implemented by 
appropriate adjustment of the amplitude of the displacement operator $\hat{D}(\cdot)$ (Fig.~\ref{fig:ProblemStatement}), i.e., $\hat{D}(-\alpha) \equiv \exp(-\alpha\hat{a}^\dagger+\alpha^*\hat{a})$ for the Kennedy receiver, where $\hat{a}^\dagger$ and $\hat{a}$ are the mode's creation and annihilation operators; $\hat{D}(-\beta)$ for the GK receiver, where $\beta>0$ is chosen to minimize $P^{(1)}_\epsilon$; and $\hat{D}(0)$ for the DD receiver.  For the six hypothesis-receiver pairs, single-copy states were detected in $1\,\upmu$s slices over a 1\,ms interval thus comprising 1000 state copies in each case.  With $\{|n_D\rangle\}$ being the photon-number states, the $\rho_{\rm coh}$ and $\rho_{\rm th}$ likelihoods for each receiver can be derived from their single-copy photon-counting statistics under the two hypotheses:  $\langle n_D|\rho_{\rm coh}|n_D\rangle$ and $\langle n_D|\rho_{\rm th}|n_D\rangle$ for the DD receiver; $\langle n_D|\hat{D}(-\alpha)\rho_{\rm coh}\hat{D}^\dagger(-\alpha)|n_D\rangle$ and $\langle n_D|\hat{D}(-\alpha)\rho_{\rm th}\hat{D}^\dagger(-\alpha)|n_D\rangle$ for the Kennedy receiver; and $\langle n_D|\hat{D}(-\beta)\rho_{\rm coh}\hat{D}^\dagger(-\beta)|n_D\rangle$ and $\langle n_D|\hat{D}(-\beta)\rho_{\rm th}\hat{D}^\dagger(-\beta)|n_D\rangle$ for the GK receiver.  Here the density operators $\rho_{\rm coh}$ and $\rho_{\rm th}$ given earlier are taken to have average photon numbers $\bar{n}_R = \eta\bar{n}_S$, where $\bar{n}_S$ is the source's average photon number and $\eta = 0.45$ is our receiver's overall efficiency (see Appendix B).  Receiver efficiency scaling preserves both coherent states and thermal states, so we shall compare our experimental results with the QCB and Helstrom bound evaluated as functions of $\bar{n}_R$.  

The Kennedy and GK receivers make hard decisions~\cite{footnote1} based on the photon counts from each of the $M$ copies, viz., for $m = 1,2,\ldots M$, they use their single-copy MLDR to decide $\rho_{\rm coh}$ or $\rho_{\rm th}$ received based on the photon counts produced by the $m$th illuminating state.  The Kennedy receiver's single-copy MLDR reduces to decide $\rho_{\rm coh}$ if no counts result from the $m$th copy and decide $\rho_{\rm th}$ otherwise, as does the GK receiver's single-copy MLDR in the photon-starved regime of interest, $\bar{n}_R \ll 1$.  These receivers' final (multi-copy) decisions are made with a binomial-distribution MLDR on the total number $n_{\rm coh}$ of $\rho_{\rm coh}$ hard decisions using, as that distribution's two possible success probabilities, $p_{\rm coh}\equiv P(\mbox{decide }\rho_{\rm coh}\mid \rho_{\rm coh}\mbox{ true})$ and $q_{\rm coh}\equiv P(\mbox{decide }\rho_{\rm coh}\mid \rho_{\rm th}\mbox{ true})$.  The GK receiver's multi-copy MLDR reduces to decide $\rho_{\rm coh}$ when $n_{\rm coh} > n^{\ast}$ and decide $\rho_{\rm th}$ otherwise, where
\begin{equation}
    n^{\ast}=M\ln\!\left[\frac{1-p_{\rm coh}}{1-q_{\rm coh}}\right]\Big/\ln\!\left[\frac{p_{\rm coh}(1-q_{\rm coh})}{q_{\rm coh}(1-p_{\rm coh})}\right].
    \label{eq:MLDRthreshold}
\end{equation}
The Kennedy receiver has $p_{\rm coh} = 1$, so its multi-copy decision is to decide $\rho_{\rm coh}$ if $n_{\rm coh} = M$ and decide $\rho_{\rm th}$ otherwise.  In the ideal case it is easily shown that the Kennedy receiver saturates the QCB:
\begin{equation}
        \xi^{\rm Ken} = \xi^{\rm QCB} = \bar{n}_R/(\bar{n}_R+1)+\ln(\bar{n}_R+1).
           \label{eq:QCB_Kennedy_Chernoff}
\end{equation} 
The DD receiver uses its single-copy likelihoods to realize the optimum multi-copy MLDR.  

\begin{figure}[bt]
	\includegraphics[width  = \columnwidth]{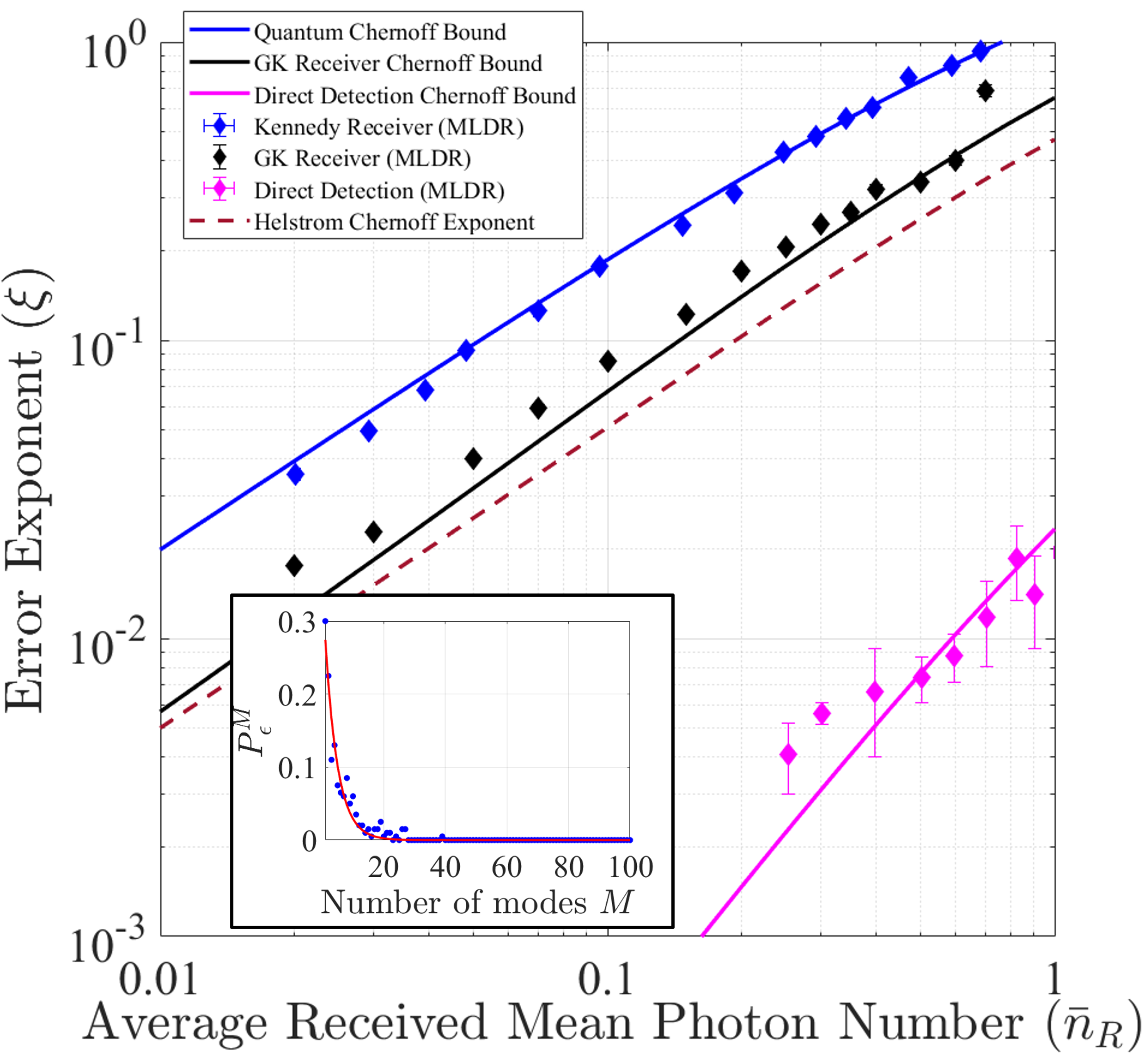}
	\caption{Experimentally measured error exponents (diamonds) and theoretically calculated Chernoff bounds to the error exponents (solid curves) versus average received photon number ($\bar{n}_R$) for multi-copy coherent-state versus thermal-state  discrimination.  Blue solid curve: quantum Chernoff bound from Eq.~\eqref{eq:QCB_Kennedy_Chernoff}. Vertical black dashed line: Kennedy receiver's 100$\times$ Chernoff-exponent advantage over the DD receiver at $\bar{n}_R = 0.6$.  Red dashed curve:  multi-copy Chernoff exponent for a receiver whose copy-by-copy decisions saturate the single-copy Helstrom bound.  (Inset) Experimental $P_\epsilon^{(M)}$ versus $M$ values (points) for the Kennedy receiver at $\bar{n}_R = 0.2$ and their experimental fit (curve) to $P_{\epsilon}^{(M)}=a\exp(-M\xi)/2$. \label{fig:ExpResults}
	}
\end{figure}

Our experimental setup had extraneous counts---a combination of stray-light and dark counts---measured to be $4\times 10^{-4}/\upmu$\,s.  Consequently, to make a decision about which state was received, we first used the photon-count data for each of the six hypothesis-receiver pairs to create photon-count histograms that we then employed, in lieu of those pairs' theoretical probability distributions, to generate the single-copy MLDRs for the Kennedy and GK receivers, and the single-copy likelihoods for the DD receiver.  The rest of the experiment followed the decision procedure described earlier.  In particular, for each $\bar{n}_R$ and each receiver, we randomly selected $M$ measurements from the 1000 measurement results available, applied the Kennedy and GK receivers' experimentally-determined single-copy MLDRs to those $M$ measurements, and then employed those hard decisions in the binomial-distribution MLDRs to obtain their multi-copy decisions.  In contrast, the DD receiver used its experimentally-determined single-copy likelihoods in its optimum multi-copy MLDR.  For each $M$ value, this process was repeated up to 100 times to estimate the conditional error probabilities, $P(\mbox{decide }\rho_{\rm coh}^{\otimes M}\vert \rho_{\rm th}^{\otimes M}\mbox{ true})$ and $P(\mbox{decide }\rho_{\rm th}^{\otimes M}\vert\rho_{\rm coh}^{\otimes M}\mbox{ true})]$, from which we computed the experimental error probability $P_{\epsilon}^{(M)}$. We then performed a least squares fit of the error probabilities for different values of $M$ to $P_{\epsilon}^{(M)} = a\exp(-M\xi)/2$ with $a$ and $\xi$ being free parameters. As an example, Fig.~\ref{fig:ExpResults}'s inset plots $P^{(M)}_{\epsilon}$ versus $M$ for the Kennedy receiver when $\bar{n}_{R} = 0.2$
along with the fit. Similarly good exponential fits were found for all three receivers over the photon-starved $\bar{n}_R$ values shown in Fig.~\ref{fig:ExpResults}. We use the fitted $\xi$'s as the experimental error exponents for the three receivers. For each receiver, data collection and computation of $P_{\epsilon}^{(M)}$ was performed 5 times to obtain an uncertainty for the reported error exponent. 
 
\subsection{Results}
Figure~\ref{fig:ExpResults} shows the Kennedy, GK, and DD receiver's multi-copy experimental error exponents versus $\bar{n}_R$ along with multi-copy theoretical results for the QCB, the Chernoff exponent ($\xi^{\rm Helstrom}$) for a receiver whose copy-by-copy decisions saturate the single-copy Helstrom bound \cite{Helstrom1964}, and the Chernoff exponents ($\xi^{\rm Ken}$, $\xi^{\rm GK}$, $\xi^{\rm DD}$) for the Kennedy, GK, and DD receivers.  Reference~\cite{habif2021quantum} showed that the GK receiver approaches Helstrom-bound performance in photon-starved single-copy discrimination between a coherent state and a thermal state. Figure~\ref{fig:ExpResults} shows that in this paper's multi-copy setting, the GK receiver fails to approach the QCB.  In contrast, the Kennedy receiver, which does not achieve the single-copy Helstrom bound~\cite{habif2021quantum}, is seen in Fig.~\ref{fig:ExpResults} to have an experimental error exponent in excellent agreement with the QCB for the multi-copy case.  
We also see that at $\bar{n}_{R} = 0.6$ the Kennedy receiver enjoys a 100$\times$ advantage in error exponent over the DD receiver.  
Finally, we note that $\xi^{\rm Ken}= \xi^{\rm QCB}$ exceeds both $\xi^{\rm GK}$ and $\xi^{\rm Helstrom}$ by at least a factor of two.  

\section{Discussion}
To make sense of the hierarchy of multi-copy receiver performance in Fig.~\ref{fig:ExpResults}, consider multi-copy discrimination between two arbitrary, equiprobable states $\rho_1^{\otimes M}$ and $\rho_2^{\otimes M}$ with a receiver that makes hard-decision measurements on each copy using a fixed, two-element, positive operator-valued measurement. In general, the multi-copy decision can be obtained using a binomial-distribution MLDR via a simple threshold test on the $M$ single-copy results, cf.~Eq.~\eqref{eq:MLDRthreshold}. In this setting, a sub-optimal single-copy error probability can be tolerated in the effort to reduce the multi-copy error, as exemplified by the single-copy versus multi-copy behavior we have just observed for the Kennedy and GK receivers.  

To probe this behavior further, we parameterize the two conditional error probabilities for single-copy discrimination with a particular measurement as $P(\mbox{decide }\rho_2\mid \rho_1\mbox{ true}) = (1+b)P^{(1)}_\epsilon$ and $P(\mbox{decide }\rho_1\mid \rho_2\mbox{ true}) = (1-b)P^{(1)}_\epsilon$, where $b \in [-1,1]$ quantifies the bias of the single-copy decision toward one type of conditional error or the other. For a symmetric measurement, i.e., with a bias of $b=0$, the multi-copy Chernoff exponent is easily found to be $\xi^{\rm Meas}=-\ln(Q_{s_{\rm min}})$, where $Q_{s_{\rm min}} = 2\sqrt{P^{(1)}_\epsilon (1-P^{(1)}_\epsilon)}$.  By series expansion (see Appendix~\ref{AppC}), we show that in the vicinity of $b=0$
\begin{equation}
		Q_{s_{\rm min}} = 2\sqrt{P_{\epsilon}^{(1)}(1-P_{\epsilon}^{(1)})}-G\big(P_{\epsilon}^{(1)}\big)b^2+O\big(b^4\big),
		\label{eq:Qsmin}
\end{equation}
where $G(x)$ is strictly positive for $x\in[0,1/2]$. Equation~\eqref{eq:Qsmin}, which holds even for two mixed states $\rho_1$ and $\rho_2$, can be used to find the multi-copy Chernoff exponent as a function of the single-copy error probability for any copy-by-copy measurement that approaches the unbiased condition.  

We now describe how quantum fidelity can be used to prove the multi-copy sub-optimality of certain copy-by-copy measurements. Let $P_{\epsilon,\rm min}^{(1)}$ be the single-copy Helstrom-bound error probability. The lower bound $P_{\epsilon,\rm min}^{(1)}\ge \left[1-\sqrt{1-F(\rho_1,\rho_2)}\right]/2$ \cite{Fuchs1999} then leads to
\begin{align}
        \xi^{\rm Meas}&\leq-\frac{1}{2}\ln\!\left[F(\rho_1,\rho_2)\right] \nonumber\\
        &+\frac{G\!\Big(\!\left[1-\sqrt{1-F(\rho_1,\rho_2)}\right]\Big/2\Big)}{\sqrt{F(\rho_1,\rho_2)}}b^2
        +O\big(b^4\big),
	\label{eq:CEBoundBSC}
\end{align}
where the bias parameter $b$ can be identified for a given measurement on a case-by-case basis. Crucially, when $\rho_1 = |\psi_1\rangle\langle\psi_1|$ and $\rho_2$ is a mixed state---as in our coherent-state versus thermal-state problem---the quantum Chernoff exponent is known to be $\xi^{\rm QCB}=-\ln\!\left[F(\rho_1,\rho_2)\right]$~\cite{kargin2005chernoff}, where the fidelity is given by $F(\rho_1,\rho_2)=\langle\psi_1|\rho_2|\psi_1\rangle$. Therefore, the inequality in \eqref{eq:CEBoundBSC} reveals that, for an arbitrary pure-state versus mixed-state discrimination task, any unbiased single-copy measurement will exhibit a Chernoff exponent that is \emph{at most} half that of the quantum Chernoff exponent, \begin{equation}
	\xi^{\rm Meas}\leq \xi^{\rm QCB}/2+O\big(b^2\big).
	\label{eq:purevsmixedgap}
\end{equation}
So, because the Helstrom measurement for coherent-state versus thermal-state discrimination is unbiased in the photon-starved  $\bar{n}_{\rm R}\to 0$ limit---see Appendix~\ref{AppD} for details---the measurement that is quantum-optimal for single-copy discrimination is strictly sub-optimal in its multi-copy Chernoff exponent $\xi^{\rm Helstrom}$ by at least a factor of two, as seen in Fig.~\ref{fig:ExpResults}. In Appendix~\ref{AppD}, we show that this gap is exactly a factor of four in the photon-starved limit~\cite{habif2018quantum}.

The inequality in \eqref{eq:CEBoundBSC} also reveals the role played by single-copy bias in the multi-copy Chernoff exponents for copy-by-copy measurements. Note that the $b^2$ terms in \eqref{eq:CEBoundBSC} and~\eqref{eq:purevsmixedgap} will be strictly positive because of the positivity of the $G(x)$ function. This means that introducing a small amount of bias in a binary measurement necessarily \emph{increases} the upper bound on the Chernoff exponent, allowing the enforced gap with respect to the quantum Chernoff exponent to possibly \emph{decrease} to less than a factor of two. We conclude that, for any pure-state versus mixed-state discrimination task, bias in the single-copy conditional error probabilities is \emph{necessary} for the Chernoff exponent of any copy-by-copy measurement to break free from two-fold sub-optimality with respect to the QCB. For coherent-state versus thermal-state discrimination, a GK measurement that makes copy-by-copy hard decisions is shown in Appendix~\ref{AppD} to exhibit a bias of $b=(\sqrt{e}-2)/\sqrt{e}\approx-0.2131$ in the photon-starved limit. Appendix~\ref{AppD} also shows that the resulting upper bound on $\xi^{\rm GK}$ from \eqref{eq:CEBoundBSC} has a gap with respect to the quantum Chernoff exponent of at least a factor of $e/\big[1+(1-\sqrt{e})^2\big]\approx1.9132$. The behavior of the GK measurement for coherent-state versus thermal-state discrimination is again confirmed to be \emph{nearly} equivalent to that of the Helstrom measurement \cite{habif2021quantum}, but its small amount of bias affords it a small decrease in its sub-optimality factor. On the other hand, the Kennedy receiver is maximally biased with $b=1$, and it achieves $\xi^{\rm Ken}=\xi^{\rm QCB}$ exactly. Indeed, it is easily seen that a maximally-biased (i.e., $|b| = 1$) single-copy receiver will achieve the QCB in multi-copy pure-state versus mixed-state discrimination.  The Chernoff exponents shown in Fig.~\ref{fig:ExpResults} illustrate the role of measurement bias as a necessary ingredient for a receiver to approach the quantum-optimal multi-copy error decay rate for pure-state versus mixed-state discrimination. 

At this point, it is worthwhile to take a step back and interpret the behavior of the Kennedy receiver in our multi-copy coherent-state versus thermal-state problem by using semiclassical---shot noise plus excess noise---photodetection theory, as was done in~\cite{Kennedy} for binary phase-shift keyed laser communications without background noise.  The Kennedy receiver's single-use hard decision for discriminating between laser light and  chaotic radiation---the classical analog of discriminating between coherent and thermal quantum states---is to decide laser-light if and only if no counts occur.  Ordinarily, in classical communications, a hard-decision receiver gives sub-optimal performance in the multi-copy case because hard-decision processing on each copy usually discards information needed for optimum multi-copy performance.  In other words, hard decisions generally do not contain all the information needed for optimum multi-copy discrimination's sufficient statistic.  In such cases a soft-decision single-copy receiver, i.e., one that preserves what is needed for optimum multi-copy discrimination's sufficient statistic, will then outperform its hard-decision counterpart. See Appendix~\ref{AppE} for a classical example involving multi-copy discrimination of a signal embedded in additive white Gaussian noise that illustrates the performance gap between hard-decision and soft-decision single-copy receivers.  In our multi-copy coherent state versus thermal state problem, however, copy-by-copy Kennedy reception \emph{does} preserve the information needed for optimum multi-copy reception's sufficient statistic, viz., whether any of the single-copy decisions were $\rho_{\rm th}$.  Thus the Kennedy receiver is a hard-decision receiver for multi-copy coherent-state versus thermal-state discrimination whose single-copy decisions are the sufficient statistic needed for optimum multi-copy performance.  

In closing, let us consider how to realize maximally-biased single-copy reception for more general pure-state versus mixed-state scenarios than the coherent-state versus mixed-state discrimination in which the Kennedy receiver suffices.  Suppose that the pure state $|\psi_1\rangle$ is a single-mode Gaussian state.  In general, such a state is a non-zero mean squeezed state.  Theoretically, it can be generated from the vacuum state $|0\rangle$ via~\cite{Weedbrook}
\begin{equation} 
|\psi_1\rangle = \hat{S}(z)\hat{D}(\alpha)|0\rangle,
\label{sqstate}
\end{equation}
where $\hat{D}(\alpha)$ is the displacement operator given earlier, and
\begin{equation}
\hat{S}(z) \equiv \exp\!\left[(z^*\hat{a}^2-z\hat{a}^{\dagger 2})/2\right], \mbox{ $z$ a complex number,}
\end{equation}
is the squeeze operator.  In practice, it can be generated by driving an optical parametric amplifier with laser light, because $\hat{D}(\alpha)|0\rangle$ is the coherent state $|\alpha\rangle$, and parametric amplification can perform the squeezing operation.  We know that $\xi^{\rm QCB} = -\ln(\langle \psi_1|\hat{\rho}_2|\psi_1\rangle)$ is the Chernoff error exponent for multi-copy discrimination between $|\psi_1\rangle$ and $\hat{\rho}_2$.  From Eq.~(\ref{sqstate}), it follows that applying the squeeze operator $\hat{S}^\dagger(z)$ to the received light---using a parametric amplifier---transforms the preceding discrimination problem into the equivalent one of choosing between $|\alpha\rangle$ and $\hat{S}^\dagger(z)\hat{\rho}_2\hat{S}(z)$, for which  copy-by-copy Kennedy reception achieves multi-copy quantum Chernoff exponent.  

The preceding analysis is readily extended to the single-copy pure state being a multi-mode squeezed state.  Such a state can be generated by driving a multi-mode parametric interaction with multi-mode coherent-state light.  Hence, undoing that multi-mode squeezing operation reduces the binary state-discrimination task to deciding between a multi-mode coherent state and a multi-mode mixed state.  By appropriate choice of spatio-temporal-polarization mode, the multi-mode coherent state in question becomes single mode, hence amenable to vacuum-or-not measurement by use of a Kennedy receiver.

We now have a multi-copy QCB-achieving receiver realization for a much broader class of problems than just the coherent-state versus thermal-state discrimination we demonstrated experimentally.  But is there a way to realize such receivers for arbitrary pure-state versus mixed-state problems?  The answer is perhaps.  The key to both the coherent-state and squeezed-state receivers is their reducing the discrimination task to a vacuum-or-not measurement by arranging that the pure state is transformed to the vacuum via a unitary operation, i.e., $\hat{D}(-\alpha)$ for the coherent-state (Kennedy) receiver and $\hat{D}(-\alpha)\hat{S}^\dagger(z)$ for the single-mode squeezed-state (parametric-amplifier-augmented Kennedy) receiver.  Photon detection in vacuum-or-not reception is then maximally biased, i.e., the pure state is detected with probability one.  Thus, if a given pure state can be generated from a unitary $\hat{U}$ via $|\psi_1\rangle = \hat{U}|\alpha\rangle$, \emph{and} we have a nonlinear interaction that recovers the coherent state $|\alpha\rangle$ by realizing $\hat{U}^\dagger|\psi_1\rangle$, then preceding a Kennedy receiver with the $\hat{U}^\dagger$ realization will yield the desired result.  Interestingly, non-destructive vacuum-or-not reception has already proven to achieve the Holevo capacity of the bosonic pure-loss channel~\cite{Wilde,footnote2}.

\begin{acknowledgements}
Research at USC was sponsored by the Army Research Office and was accomplished under Grant Number W911NF-20-1-0235. MG and SG acknowledge support from the DARPA IAMBIC Program under Contract Number HR00112090128. JHS acknowledges support from the MITRE Corporation's Quantum Moonshot Program.  The views and conclusions contained in this document are those of the authors and should not be interpreted as representing the official policies, either expressed or implied, of the Army Research Office, DARPA or the U.S. Government. The U.S. Government is authorized to reproduce and distribute reprints for Government purposes notwithstanding any copyright notation herein.
\end{acknowledgements}

\appendix

\section{The quasi photon-number resolving detector \label{AppA}}
We did not have a true photon-number resolving detector for our experiment.  Instead, because the photon flux in our photon-starved operating regime was sufficiently low, we were able to achieve quasi photon-number resolving (qPNR) detection by post-processing measurements made within 1\,$\upmu$s time windows using a Perkin-Elmer Si avalanche photodiode single-photon detector (SPCM-AQR-14).  This detector's minimum detection time---set by its dead time---is 50\,ns.  If the probability that more than one photon illuminates the detector within a 50\,ns interval is sufficiently low, compared to that for single-photon illumination, then each detection event can be regarded as a projection onto the Fock state $|1\rangle$.  The number of detection events within the 1\,$\upmu$s measurement window then constitutes qPNR operation.

\section{Impact of sub-unit receiver efficiency on $\xi^{\rm Ken}$ \label{AppB}}
As stated in the main text, our receiver implementation suffered optical losses  attributed to non-idealities in the optical paths as well as the sub-unit quantum efficiency of its Si single-photon detector.  These effects resulted in an overall receiver efficiency $\eta = 0.45$, which is why the main text's comparisons between theory and experiment were referenced to the average received photon number $\bar{n}_R \equiv \eta \bar{n}_S$, where $\bar{n}_S$ our source's average photon number.  Here we quantify the impact of receiver efficiency on the Kennedy receiver's error exponent.

\begin{figure}[h!]
    \centering
    \includegraphics[width  = \columnwidth]{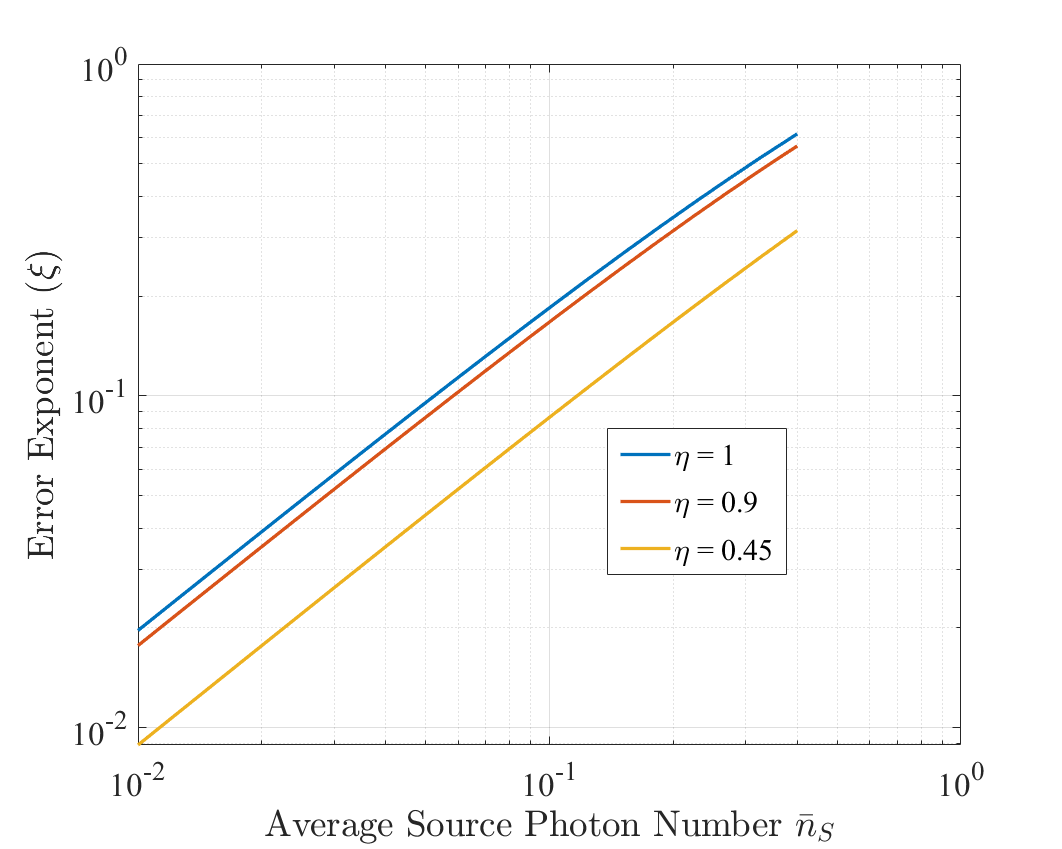}
    \caption{The error exponent achieved by the Kennedy receiver for three different values of receiver efficiency $\eta$. When $\eta = 1$ the Kennedy receiver is ideal and achieves the quantum Chernoff exponent.}
    \label{fig:KenEta}
\end{figure}

Figure \ref{fig:KenEta} plots the Kennedy receiver's error exponent, $\xi^{\rm Ken}$, versus $\bar{n}_S$ for: ideal reception, $\eta = 1$; the $\eta = 0.45$ achieved by our equipment; and the $\eta = 0.9$ that could be achieved with a state-of-the-art superconducting nanowire single-photon detector~\cite{Reddy:20,Chang2021detecting}.

\section{Asymptotic Multi-Copy Sub-Optimality of Optimal Copy-By-Copy Measurements \label{AppC}}
Consider multi-copy ($M$-ary) discrimination between two arbitrary, equiprobable states  $\rho_1^{\otimes M}$ and $\rho_2^{\otimes M}$ with a receiver that makes hard-decision measurements on each copy using a fixed two-element positive operator-valued  measurement (POVM), and then obtains its multi-copy decision by applying a decision rule to the POVM outcomes. Here we prove the conditions under which a fundamental gap exists between this receiver's asymptotic error exponent and that of the optimum multi-copy receiver, i.e., the quantum Chernoff exponent.  

Let the single-copy conditional error probabilities be $p \equiv P(\mbox{decide }\rho_2\vert\rho_1\mbox{ true})$ and $q \equiv P(\mbox{decide }\rho_1\vert\rho_2\mbox{ true})$, so that the single-copy error probability is $P_{\epsilon}^{(1)}=(p+q)/2$. The classical Chernoff exponent for such a measurement is given by $\xi^{\rm Meas}=-\ln(Q_{s_{\rm min}})$, where $Q_{s_{\rm min}}=\min_{0\leq s\leq 1}Q_s$ and \cite{VanTrees2013}
\begin{equation}
    Q_s=(1-p)^sq^{1-s}+p^s(1-q)^{1-s}.
    \label{eq:Qs}
\end{equation}
Because $Q_s$ is strictly convex \cite{Audenaert}, it has a unique minimum, which occurs at 
\begin{equation}
	s_{\rm min}=\frac{\ln\!\big[(1-q)\ln[(1-q)/p]\big]-\ln\!\big[q\ln[(1-p)/q]\big]}{\ln\!\Big[(1-p)(1-q)/pq\Big]}.
	\label{eq:smin}
\end{equation}
We now parameterize the two single-copy error probabilities by $p=(1+b)P_{\epsilon}^{(1)}$ and $q=(1-b)P_{\epsilon}^{(1)}$, where $b\in[-1,1]$ quantifies the bias of the single-copy decision toward one type of conditional error or the other.  Expanding $Q_{s_{\rm min}}$ around $b=0$, yields the main text's Eq.~(3) with
\begin{widetext}
\begin{equation}
		G\big(P_{\epsilon}^{(1)}\big)=\exp\!\Bigg(\frac{\ln(\bar{P}_{\epsilon})^2-\ln(P_{\epsilon}^{(1)})^2}{2\big[\ln(\bar{P}_{\epsilon})-\ln(P_{\epsilon}^{(1)})\big]}\Bigg)\!\left(\frac{\bar{\bar{P}}_{\epsilon}^2+2\tanh^{-1}(\bar{\bar{P}}_{\epsilon})\big[\big(\bar{\bar{P}}_{\epsilon}+2(P_{\epsilon}^{(1)})^2\big)\tanh^{-1}(\bar{\bar{P}}_{\epsilon})-\bar{\bar{P}}_{\epsilon}\big]}{\bar{P}_{\epsilon}^2\ln\big(P_{\epsilon}^{(1)}/\bar{P}_{\epsilon}\big)}\right).
	\label{eq:G}
\end{equation}
\end{widetext}
Here, $\bar{P}_{\epsilon} \equiv 1-P_{\epsilon}^{(1)}$, $\bar{\bar{P}}_{\epsilon}\equiv 1-2P_{\epsilon}^{(1)}$, and $G\big(P_{\epsilon}^{(1)}\big)$ 
is strictly positive for $P_{\epsilon}^{(1)}\in[0,1/2]$. 

Next, we write the series of inequalities
\begin{align}
		Q_{s_{\rm min}} \geq&  2\sqrt{P_{\epsilon,\rm min}^{(1)}\left(1-P_{\epsilon,\rm min}^{(1)}\right)}-G\big(P_{\epsilon,\rm min}^{(1)}\big)b^2+O\big(b^4\big) \nonumber \\
		\geq& \sqrt{F(\rho_1,\rho_2)}-G\bigg(\Big[1-\sqrt{1-F(\rho_1,\rho_2)}\Big]\Big/2\bigg)b^2 \nonumber \\
		&+O\big(b^4\big).
	\label{eq:QsInequalities}
\end{align}
The first inequality is the Helstrom bound $P_{\epsilon}^{(1)}\geq P_{\epsilon, \rm min}^{(1)} =\big(1-\norm{\rho_1-\rho_2}_1/2\big)/2$ \cite{Helstrom1964}, where $\norm{\cdot}_1$ is the operator trace norm, and the second comes from the well-known result of Fuchs and van de Graaf \cite{Fuchs1999},
\begin{equation}
	1-\sqrt{F(\rho_1,\rho_2)}\leq \norm{\rho_1-\rho_2}_1\leq \sqrt{1-F(\rho_1,\rho_2)},
	\label{eq:FuchsVanDeGraff}
\end{equation}
with $F(\rho_1,\rho_2) \equiv \Tr\big[\sqrt{\sqrt{\rho_1}\rho_2\sqrt{\rho_1}}\big]^2$ being the quantum fidelity between $\rho_1$ and $\rho_2$. Monotonicity is also used in both inequalities in \eqref{eq:QsInequalities}, as $\sqrt{x(1-x)}$ increases monotonically with increasing $x\in[0,1/2]$, and $G(x)$ decreases monotonically with increasing $x\in[0.1166,1/2]$. Taylor expanding the logarithm in $\xi^{\rm Meas} \equiv -\ln(Q_{s_{\rm min}})$, we get the main text's (4).  That inequality thus places an upper bound on the Chernoff exponent for \emph{all} multi-copy, binary quantum-state discrimination tasks that use identical, hard-decision, copy-by-copy measurements and any decision rule.  
We can also lower bound the Chernoff exponent for the Helstrom measurement, which always consists of a two-element POVM and attains the fundamental minimum single-copy error probability $P_{\epsilon,\rm min}^{(1)}$ \cite{Helstrom1964}. Specifically, using the lower bound from \eqref{eq:FuchsVanDeGraff}, we find
\begin{eqnarray}
		\xi^{\rm Helstrom}\geq&& -\ln\!\Big[\sqrt{F(\rho_1,\rho_2)}\left(2-\sqrt{F(\rho_1,\rho_2)}\right)\Big]\Big/2 \nonumber \\
		&&+O\big(b^2\big).
	\label{CEHelstromBoundBSC}
\end{eqnarray}

We now compare the classical Chernoff exponent for the Helstrom measurement to the quantum Chernoff exponent in three general scenarios that span all binary quantum hypothesis tests. In all three cases, the first inequality in \eqref{eq:QsInequalities} is satisfied with an equality because the Helstrom measurement always saturates the Helstrom bound \cite{Helstrom1964}. First, when $\rho_1=\outerproduct{\psi_1}{\psi_1}$ and $\rho_2=\outerproduct{\psi_2}{\psi_2}$, the quantum Chernoff exponent is exactly given by \cite{kargin2005chernoff}
\begin{equation}
	\xi^{\rm QCB}=-\ln\!\big[F(\rho_1,\rho_2)\big],
	\label{eq:QCB_pure}
\end{equation}
where $F(\rho_1,\rho_2)=\abs{\innerproduct{\psi_1}{\psi_2}}^2$. For this pure-state case, the Helstrom measurement is comprised of linear projectors onto $\ket{v_1} \equiv (\ket{0}+\ket{1})/\sqrt{2}$ and $\ket{v_2} \equiv (\ket{0}-\ket{1})/\sqrt{2}$, where $|0\rangle \equiv (|\psi_1\rangle + e^{-i\phi}|\psi_2\rangle)/\sqrt{2(1+|\langle \psi_1|\psi_2\rangle|)}$ and  $|1\rangle \equiv (|\psi_1\rangle - e^{-i\phi}|\psi_2\rangle)/\sqrt{2(1-|\langle \psi_1|\psi_2\rangle|)}$, with $e^{i\phi} \equiv \langle \psi_1|\psi_2\rangle/|\langle \psi_1|\psi_2\rangle|$, is an orthonormal basis for the Hilbert space spanned by $|\psi_1\rangle$ and $|\psi_2\rangle$.  It is then simple to show that the conditional error probabilities  $p=\abs{\innerproduct{v_2}{\psi_1}}^2$ and $q=\abs{\innerproduct{v_1}{\psi_2}}^2$ satisfy $p=q=\Big[1-\sqrt{1-F(\rho_1,\rho_2)}\Big]/2$, hence the pure-state Helstrom measurement is always unbiased, i.e., it has $b=0$. From this observation, along with the fact that pure states saturate the upper bound in \eqref{eq:FuchsVanDeGraff}, the second inequality in \eqref{eq:QsInequalities} becomes an equality and all terms but the first on the right hand side vanish. We are left with $\xi^{\rm Helstrom}=-\ln\!\big[F(\rho_1,\rho_2)\big]/2$, confirming that the Helstrom measurement, which is quantum-optimal for single-copy discrimination, achieves an asymptotic error exponent that is sub-optimal by \emph{exactly} a factor of two compared to the asymptotic quantum limit for multi-copy pure-state versus pure-state discrimination~\cite{Acin2005}.

When $\rho_1=\outerproduct{\psi_1}{\psi_1}$  and $\rho_2$ is a mixed state, the quantum Chernoff exponent is again given by Eq.~\eqref{eq:QCB_pure}, where now $F(\rho_1,\rho_2)=\abs{\mel{\psi_1}{\rho_2}{\psi_1}}$ \cite{kargin2005chernoff}. The general description of the Helstrom measurement involves the eigenspectrum of  $\Lambda=\rho_1-\rho_2$, viz., the POVM elements $\Pi^{\rm Helstrom}_1$ and $\Pi^{\rm Helstrom}_2$ project, respectively, onto the subspaces associated with $\Lambda$'s non-negative and negative eigenvalues~\cite{Helstrom1964}. In general, the resulting conditional error probabilities, $p=\Tr[\Pi^{\rm Helstrom}_2\rho_1]$ and $q=\Tr[\Pi^{\rm Helstrom}_1\rho_2]$, can be biased ($b \neq 0$), although we will show that they are approximately unbiased for the main text's photon-starved discrimination task. Still, the second inequality in \eqref{eq:QsInequalities} leads to (4) in the main text. From Eq.~\eqref{eq:QCB_pure}---which holds when at least one state is pure---we then conclude that any measurement that is unbiased to first order in $b$ exhibits an asymptotic error exponent that is \emph{at least} a factor of two lower than the quantum Chernoff exponent for pure-state versus mixed-state hypothesis testing, as in the main text's (5).

When $\rho_1$ and $\rho_2$ are both mixed states, the main text's (4) still holds. 
Now,  the quantum Chernoff exponent cannot be determined from the fidelity, but it is known to obey \cite{kargin2005chernoff}
\begin{equation}
	-\ln\!\big[F(\rho_1,\rho_2)\big]/2 \leq \xi^{\rm QCB} \leq -\ln\!\big[F(\rho_1,\rho_2)\big].
	\label{eq:QCB_mixed}
\end{equation}
To first order in $b$, the bounded intervals for $\xi^{\rm Meas}$ and $\xi^{\rm QCB}$ intersect at exactly the point $-\ln\!\big[F(\rho_1,\rho_2)\big]/2$, implying that we cannot show a gap between the quantum and classical Chernoff exponents; rather, we simply reaffirm that $\xi^{\rm Meas}\leq \xi^{\rm QCB}$, which is the operational meaning of the quantum Chernoff bound \cite{Audenaert,Nussbaum2009}. It is possible that restricting the states further, e.g., by considering only Gaussian states \cite{Banchi2015}, or states that are separated by a vanishingly-small operator perturbation \cite{Grace2021}, could lead to a tighter lower bound on the quantum Chernoff exponent and thus reveal a gap between unbiased measurements and the quantum limit. It is interesting to note that for a specific multi-copy binary hypothesis test involving two-mode zero-mean Gaussian states under each hypothesis that emerges in the analysis of a quantum radar for target-detection, the Chernoff exponent attained by the copy-by-copy Helstrom measurement (followed by a majority-vote decision rule) is a factor of two lower than the quantum Chernoff exponent~\cite{PhysRevA.80.052310}. We leave a full investigation of this topic for future work. 

\section{Asymptotic Receiver Performance for Coherent-State versus Thermal-State Discrimination \label{AppD}}

In this section we analytically investigate the single-copy-optimized and multi-copy-optimized asymptotic error exponents for multi-copy discrimination between $\rho_{\rm coh}^{\otimes M}$ and $\rho_{\rm th}^{\otimes M}$, where $\rho_{\rm coh} = |\sqrt{\bar{n}_R}\rangle\langle\sqrt{\bar{n}_R}|$ is the coherent-state density operator, and $\rho_{\rm th} = \sum_{n=0}^\infty\frac{\bar{n}_R^{n}}{(\bar{n}_R+1)^{n+1}} |n\rangle \langle n|$, with $\{|n\rangle\}$ being the state space $\mathcal{H}$'s photon-number basis, is the thermal-state density operator.  As in the main text, we will presume photon-starved single-copy operation, i.e., $\bar{n}_R \ll 1$.  The single-copy Kennedy measurement is characterized by POVM elements $\Pi^{\rm Ken}_{\rm th}=\mathcal{I}-\rho_{\rm coh}$ and $\Pi^{\rm Ken}_{\rm coh}=\rho_{\rm coh}$, where $\mathcal{I}$ is $\mathcal{H}$'s identity operator. For this measurement we have that $p \equiv P(\mbox{decide }\rho_{\rm coh}\mid \rho_{\rm th}\mbox{ true}) = \langle\sqrt{\bar{n}_R}|\rho_{\rm th}|\sqrt{\bar{n}_R}\rangle$ and $q \equiv P(\mbox{decide }\rho_{\rm th}\mid \rho_{\rm coh}\mbox{ true}) = 0$, implying that $b=1$. Because $\rho_{\rm coh}$ is a pure state, the Chernoff exponent $\xi^{\rm Ken}$ is guaranteed to equal the quantum Chernoff exponent $\xi^{\rm QCB}=-\ln\!\big[F(\rho_{\rm coh},\rho_{\rm th})\big]$ \cite{kargin2005chernoff}. The fidelity between the two states--- easily calculated in the Fock basis---is $F(\rho_{\rm coh},\rho_{\rm th})=\Tr[\rho_{\rm coh}\rho_{\rm th}]=(1+\bar{n}_R)^{-1}\exp\!\big[-\bar{n}_R/(1+\bar{n}_R)\big]$, yielding the main text's Eq.~(2) and its photon-starved behavior
\begin{equation}
	\xi^{\rm Ken}=\xi^{\rm QCB}=2\bar{n}_{\rm R}+O\big(\bar{n}_{\rm R}^2\big)
	.
	\label{eq:QCE_CEKennedy}
\end{equation}  
On the other hand, when $P_{\epsilon,\rm min}^{(1)}>0.1166$, which will be true when $\bar{n}_{\rm R} \ll 1$, we can use the fidelity between $\rho_{\rm coh}$ and $\rho_{\rm th}$ along with \eqref{eq:QsInequalities} and the first-order expansion $\ln(1+x)=x+O(x^2)$ to upper bound the Chernoff exponent for any copy-by-copy measurement as follows,
\begin{equation}
	\xi^{\rm Meas}\leq\Big[1+b^2+O\big(b^4\big)\Big]\!\Big[\bar{n}_R+O\big(\bar{n}_R^{3/2}\big)\Big].
	\label{eq:CE_UpperBound}
\end{equation}

To analyze the Helstrom measurement in the photon-starved regime, we approximate $\rho_{\rm coh}$ and $\rho_{\rm th}$ as qubit density operators by truncating them to $\mathcal{H}$'s two-dimensional subspace spanned by the vacuum and single-photon Fock states.  On this subspace, the Helstrom measurement projectors $\Pi^{\rm Helstrom}_{\rm coh}$ and $\Pi^{\rm Helstrom}_{\rm th}$ yield the conditional error probabilities $p=1/2+O\big(\bar{n}_R\big)$ and $q=1/2-\sqrt{\bar{n}_R}+O\big(\bar{n}_R\big)$. The resulting bias is $b=\sqrt{\bar{n}_R}+O\big(\bar{n}_R\big)$, showing that the Helstrom measurement for coherent-state versus thermal-state discrimination is only weakly biased in the photon-starved regime. Using \eqref{eq:CE_UpperBound}, the Chernoff exponent for the Helstrom measurement must obey  
\begin{equation}
	\xi^{\rm Helstrom}\leq \bar{n}_R+O(\bar{n}_R^{3/2}),
	\label{eq:CE_HelstromBound}
\end{equation}
enforcing a sub-optimality of at least a factor of two compared with the quantum Chernoff exponent. Using the conditional error probabilities $p$ and $q$, we find that $Q_s=1-2s(1-s)\bar{n}_R+O\big(\bar{n}_R^2\big)$.  So, to first order in $\bar{n}_R$, we have $s_{\rm min}=1/2$, and we find that the Helstrom measurement's Chernoff exponent satisfies
\begin{equation}
	\xi^{\rm Helstrom}=\bar{n}_R/2+O\big(\bar{n}_{\rm R}^2\big),
	\label{eq:CE_Helsrom}
\end{equation}
revealing that its sub-optimality, for photon-starved coherent-state versus thermal-state discrimination, is actually a factor of four. 

In the photon-starved regime, the GK measurement's POVM elements, $\Pi_{\rm coh}^{\rm GK}$ and $\Pi_{\rm th}^{\rm GK}$, correspond to detecting 0 photons or detecting at least 1 photon, respectively, after the $\hat{D}(-\beta)$ displacement, where $\beta>0$. The resulting conditional error probabilities are $p \equiv P(\mbox{decide }\rho_{\rm th}\mid \rho_{\rm coh}\mbox{ true}) = 1-\exp\!\big[-(\sqrt{\bar{n}_R}-\beta)^2\big]$ and $q \equiv P(\mbox{decide }\rho_{\rm coh}\mid \rho_{\rm th}\mbox{ true}) = (1+\bar{n}_{\rm R})^{-1}\exp\!\big[-\beta^2/(1+\bar{n}_{\rm R})\big]$~\cite{Gagliardi1995}. We found that the displacement that minimizes  $P_{\epsilon}^{(1)}$ approaches $\beta=1/\sqrt{2}$ as $\bar{n}_{\rm R}\to0$, so we used this value in our analysis of photon-starved GK reception. The conditional error probabilities then become $p=1/\sqrt{e}-\bar{n}_R/(2\sqrt{e})+O\big(\bar{n}_R^{3/2}\big)$ and $q=1-1/\sqrt{e}-\sqrt{2\bar{n}_R/e}+O\big(\bar{n}_R^{3/2}\big)$, resulting in a nonzero bias given by $b=(\sqrt{e}-2)/\sqrt{e}-2\sqrt{2\bar{n}_R}/e+O(\bar{n}_R)$. We can use \eqref{eq:CE_UpperBound} to find that the Chernoff exponent of the photon-starved GK reception is bounded from above by 
\begin{equation}
	\xi^{\rm GK}\leq\frac{2\big[1+(1-\sqrt{e})^2\big]}{e}\bar{n}_R+O(\bar{n}_R^{3/2}).
\end{equation}
Therefore, the error-exponent gap between the optimal multi-copy measurement and the copy-by-copy GK receiver must be at least a factor of $e/\big[1+(1-\sqrt{e})^2\big]\approx1.9132$ in the photon-starved limit. This gap is almost identical to the bound we found in \eqref{eq:CE_HelstromBound} for the Helstrom measurement, but we find that the bias in the GK measurement allows for the gap to be less than a factor of two. 

Finally, we consider the optimal multi-copy direct-detection (DD) receiver, i.e., one that uses photon-number resolving detection of each copy, retaining the resulting counts for use in the multi-copy maximum-likelihood decision rule. Directly computing the Chernoff exponent \cite{VanTrees2013}, we find that $Q_s=1-\big(1-2^{-s}+s/2\big)\bar{n}_R^2+O\big(\bar{n}_R^3\big)$ in the photon-starved regime.  To lowest order in $\bar{n}_R$, this expression has its minimum at $s_{\rm min}=\log_2\!\big[\ln(4)\big]=0.4712$. The resulting Chernoff exponent is then
\begin{equation}
	\xi^{\rm DD}=\frac{1-\log_2[\ln(2^e)]}{2}\bar{n}_{\rm R}^2+O\big(\bar{n}_{\rm R}^3\big),
	\label{eq:CEdirect}
\end{equation}
where the lowest-order term evaluates to $0.0430\,\bar{n}_R^2$. Table~\ref{tab:CEs} summarizes the Chernoff-exponent calculations for our Kennedy, GK, and DD receivers.

\begin{table}[h!]
	\centering
	\begin{tabular}{ |c|c| } 
		\hline
		Measurement & Chernoff Exponent ($\xi^{\rm Meas}$)\\ 
		\hline
		\rule{0pt}{3ex} QCB/Kennedy & $2\bar{n}_R+O(\bar{n}_R^2)$ \\
		\rule{0pt}{3ex} Helstrom (upper bound)& $\bar{n}_R+O(\bar{n}_R^{3/2})$ \\
		\rule{0pt}{3ex} Helstrom & $\bar{n}_R/2+O(\bar{n}_R^{2})$ \\
		\rule{0pt}{3ex} GK (upper bound)& $2\big[1+(1-\sqrt{e})^2\big]\bar{n}_R/e+O(\bar{n}_R^{3/2})$ \\
		\rule{0pt}{3ex} DD & $(1-\log_2[\ln(2^e)])\bar{n}_R^2/2+O\big(\bar{n}_ R^3\big)$ \\
		\hline
	\end{tabular}
	\caption{Analytical Chernoff exponents and bounds for photon-starved, $M$-copy, coherent-state versus thermal-state discrimination evaluated to lowest nonzero order in the average photon number $\bar{n}_R$. }
	\label{tab:CEs}
\end{table}

\section{Soft-decision versus hard-decision reception of a multi-copy signal embedded in white Gaussian noise \label{AppE}}
To illustrate the performance gap between soft-decision and hard-decision multi-copy discrimination, consider the following example from classical communication.  Suppose that a communication receiver collects the $M$ independent, identically distributed waveforms
\begin{equation}
r_m(t) = \left\{\begin{array}{ll}\sqrt{E/T} + w(t), & \mbox{under hypothesis $H_1$},\\[.05in]
w(t), &\mbox{under hypothesis $H_2$},\end{array}\right.
\end{equation}
for $t\in\mathcal{T}_m, m = 1,2,\ldots, M$.  Here: $E$ is the single-copy transmitted energy under hypothesis $H_1$; ${\mathcal{T}_m}$ is a collection of non-overlapping duration-$T$ time intervals; and $w(t)$ is zero-mean white Gaussian noise with spectral density $\sigma^2$.  Assuming the two hypotheses to be equally likely, the error probability of optimum multi-copy reception is~\cite{VanTrees2013} $P_\epsilon^{(M)} = Q(\sqrt{ME/4\sigma^2})$,
where
\begin{equation}
Q(x) \equiv \int_x^\infty\!{\rm d}y\,e^{-y^2/2}/\sqrt{2\pi}.
\end{equation}
Using the exponentially-tight upper bound~\cite{VanTrees2013}, $Q(x) \le e^{-x^2/2}/2$ for $x\ge 0$, we see that the multi-copy Chernoff exponent for this problem, viz., its soft-decision Chernoff exponent, is 
\begin{equation}
\xi^{\rm soft} = E/8\sigma^2.  
\label{xiSoft}
\end{equation}

For optimum single-copy reception, we have that the conditional error probabilities are
$p \equiv P(\mbox{decide $H_2$}\mid \mbox{$H_1$ true}) = Q(\sqrt{E/4\sigma^2})$ and 
$q \equiv P(\mbox{decide $H_1$}\mid \mbox{$H_2$ true}) = Q(\sqrt{E/4\sigma^2})$, showing that this problem is unbiased, viz., $b=0$.  The multi-copy error probability of copy-by-copy, hard-decision reception therefore has Chernoff bound
\begin{equation}
P_\epsilon^{(M)} \le \frac{\left[2\sqrt{P_\epsilon^{(1)}(1-P_\epsilon^{(1)})}\right]^M}{2},   
\end{equation}
where $P_\epsilon^{(1)} = (p+q)/2 = Q(\sqrt{E/4\sigma^2})$.    In energy-starved operation, i.e., when $E/8\sigma^2 \ll 1$ so that $P_\epsilon^{(1)}$ is close to 1/2, we have that
\begin{eqnarray}
Q(\sqrt{E/4\sigma^2})  &&= 1/2-\int_0^{\sqrt{E/4\sigma^2}}\!{\rm d}y\,e^{-y^2/2}/\sqrt{2\pi}\nonumber\\
&&\approx 1/2 - \sqrt{E/8\pi\sigma^2},
\end{eqnarray}
implying that
\begin{eqnarray}
\xi^{\rm hard} &&= -\ln\!\left[2\sqrt{Q(\sqrt{E/4\sigma^2})[1-Q(\sqrt{E/4\sigma^2})]}\right]\nonumber\\
&&\approx -\ln\!\left(\sqrt{1-E/2\pi\sigma^2}\right)\approx E/4\pi\sigma^2.
\label{xiHard}
\end{eqnarray}
From Eqs.~\eqref{xiSoft} and \eqref{xiHard} we get
\begin{equation}
\xi^{\rm soft}/\xi^{\rm hard} = \pi/2,
\end{equation}
for the error-exponent advantage of soft-decision reception over hard-decision reception.


\begin{thebibliography}{99}
\bibitem{footnote1}In multi-copy classical communications, a hard-decision receiver for $M$ independent identically distributed repetitions of a binary hypothesis test ($H_1^{\otimes M}$ versus $H_2^{\otimes M}$) does copy-by-copy MLDR decisions between the two hypotheses.  It then makes its final (multi-copy) decision using a binomial-distribution MLDR on the total number of $H_1$ hard decisions using $p_1\equiv P(\mbox{decide }H_1\mid H_1\mbox{ true})$ and $q_1\equiv P(\mbox{decide }H_1\mid H_2\mbox{ true})$ as that distribution's two possible success probabilities.
\end{thebibliography}
\end{document}